# Ultrafast nanocomposite scintillators based on Cd-enhanced CsPbCl$_3$ nanocrystals in polymer matrix

Andrea Erroi[1], Francesco Carulli[1], Francesca Cova[1], Isabel Frank[2,7], Matteo L. Zaffalon[1], Jordi Llusar[3], Sara Mecca[1], Alessia Cemmi[4], Ilaria Di Sarcina[4], Francesca Rossi[5], Luca Beverina[1], Francesco Meinardi[1], Ivan Infante[3,6], Etiennette Auffray[2] and Sergio Brovelli[1]*

[1] *Dipartimento di Scienza dei Materiali, Università degli Studi Milano - Bicocca, via R. Cozzi 55, 20125 Milano, Italy.*
[2] *CERN, Esplanade des Particules 1, 1211 Meyrin, Switzerland.*
[3] *BCMaterials, Basque Center for Materials, Applications, and Nanostructures, UPV/EHU Science Park, Leioa 48940, Spain*
[4] *ENEA Fusion and Technology for Nuclear Safety and Security Department, Casaccia R.C., Via Anguillarese 301, 00123 Rome, Italy.*
[5] *IMEM-CNR Institute, Parco Area delle Scienze 37/A, 43124 Parma, Italy.*
[6] *Ikerbasque Basque Foundation for Science, Bilbao 48009, Spain*
[7] *Ludwig Maximilian University, 80539 Munich, Germany*
*sergio.brovelli@unimib.it

## ABSTRACT

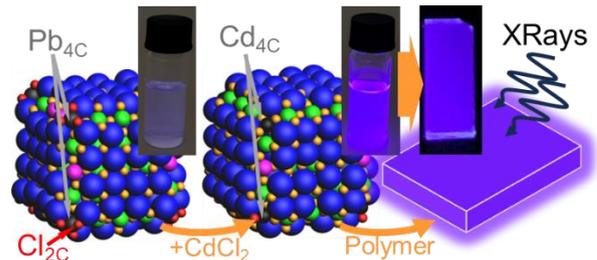

Lead halide perovskite nanocrystals (LHP-NCs) embedded in polymer matrices are gaining traction for next-generation radiation detectors. While progress has been made on green-emitting CsPbBr$_3$ NCs, scant attention has been given to the scintillation properties of CsPbCl$_3$ NCs, which emit size-tunable UV-blue light matching the peak efficiency of ultrafast photodetectors. In this study, we explore the scintillation characteristics of CsPbCl$_3$ NCs produced through a scalable method and treated with CdCl$_2$. Spectroscopic, radiometric and theoretical analysis on both untreated and treated NCs uncover deep hole trap states due to surface undercoordinated chloride ions, eliminated by Pb to Cd substitution. This yields near-perfect efficiency and resistance to polyacrylate mass-polymerization. Radiation hardness tests demonstrate stability to high gamma doses while time-resolved experiments reveal ultrafast radioluminescence with an average lifetime as short as 210 ps. These findings enhance our comprehension of LHP NCs' scintillation properties, positioning CsPbCl$_3$ as a promising alternative to conventional fast scintillators.

Scintillator materials play a pivotal role in a wide array of applications, ranging from radiation detection in the fields of nuclear physics[1], medical imaging[2-3], and homeland security[4-5], to fundamental research in high-energy physics (HEP)[6]. As the demand for efficient and versatile scintillators continues to grow, the quest for materials that exhibit superior performance characteristics intensifies[7-8]. Nanocomposite scintillators[9-12], a recent and promising development in this domain, have garnered significant attention for their potential to overcome the limitations of traditional scintillator crystals, which, although highly effective, face inherent limitations in terms of upscaling, cost-effectiveness and timing performance[1]. Conversely, commercial plastic scintillators[13], while cost-effective, versatile and fast, exhibit limited performance in terms of energy resolution and radiation hardness due respectively to their low density and inherent fragility of conjugated molecular emitters to high energy radiation[14]. To address these challenges, the integration of colloidal scintillator nanocrystals (NCs) of high atomic number elements into nanocomposite plastic matrices[10, 15] offers an innovative solution that harnesses the advantages of both worlds and has the potential to overcome the upscaling limitations of conventional scintillator crystals via cost-effective chemical means while simultaneously enhancing the performance of dye-based plastic scintillators[15].

In this framework, lead halide perovskite NCs (LHP-NCs) have rapidly garnered particular interest[9-10, 12, 16-18] due to their high emission yield[19], radiation hardness[19-20], defect tolerance[21] and unmatched scalability via room temperature synthesis methods[22]. This unique combination of chemical and physical advantages has driven a large number of studies encompassing X-ray imaging[23-26], fast timing[12, 17, 20, 27-28] and therapy[29] applications as well as the detection of gamma rays and neutrons[10, 30-31]. Despite substantial advancements, as of today, the near totality of studies in this area has focused on green-emitting $CsPbBr_3$ NCs[10, 12, 19-20, 23], with fewer investigations dedicated to their red-emitting iodine-based counterparts (mostly as single crystals or films for gamma detection)[32-35]. Most surprisingly, there has been no study to date that has addressed the scintillation properties of $CsPbCl_3$ NCs, which feature size tunable ultrafast emission in the UV-blue[36-37] and would thereby extend the spectral tunability of LHP-based nanocomposite scintillators to the typical spectral region of molecular scintillators such as 1,4-bis(5-phenyloxazol-2-yl) benzene (POPOP) ($\lambda_{EM}$~410 nm) and para-terphenyl ($\lambda_{EM}$~350 nm)[38] that match the peak efficiency of bialkali photodetectors widely used in HEP experiments (e.g. Hamamatsu R9880U-210, see **Supporting Figure S1**)[39-40]. Further $CsPbCl_3$ NCs have the potential to serve as high-Z sensitizers[41] for secondary molecular emitters in the green spectral region that typically suffer less radiation damage than the more energetic blue emitting counterparts.

To achieve a sub-100-ps time resolution, mechanisms involving the production of prompt photons, like Cherenkov emission and cross luminescence, were so far considered as the most promising solutions, overcoming the limitations of traditional UV-emitting scintillators, where the fast light signal comes from optically allowed transitions of rare-earth ions used as activators, or from strongly quenched intrinsic luminescence, as for $PbWO_4$. Indeed, rare earth-doped halides are prized for their high density and for a LY of several tens of thousands ph MeV$^{-1}$ achieved by Ce doping[42] but their scintillation lifetime cannot be shorter than a few tens of ns with Nd, or Pr activators[43-45]. On the other hand, the number of prompt Cherenkov photons from the recoil electrons resulting from a 511-keV γ interaction is very small, of the order of 20 photons per event in crystals, such as $Lu_2SiO_5$:Ce, $Lu_3Al_5O_{12}$:Ce, $Bi_4Ge_3O_{12}$ (BGO) and $PbWO_4$[46-49], thus limiting the application of Cherenkov radiators in combination with highly emissive scintillators in dual readout approaches in HEP experiments[50-52] and time of flight positron emission tomography (TOF-PET)[53-54]. Only recently, a coincidence time resolution (CTR) of ~30 ps was reported using 3.2-mm thick Cherenkov emitting crystals and applying depth-of-interaction correction algorithms[55-56]. In stark contrast, cross luminescent materials like Ba-doped[57] and Cs-based compounds[58-60] feature a high scintillation yield and a reasonably fast sub-ns emission (600 ps for $BaF_2$[61]) from the core-valence transition which however occurs in the vacuum UV spectral range (< 250 nm for $BaF_2$), where the optical transmittance and the photodetector quantum efficiency are generally low. In addition, $BaF_2$ is affected by a dominant slow luminescence contribution of around 600 ns[61-62] from self-trapped excitons, undesirable for fast emitters[63]. Other novel cross luminescent compounds showed superior performances than traditional $BaF_2$. $Cs_2ZnCl_4$ is non-hygroscopic, has a single-component decay time of 1.7 ns, longer emission wavelength and a light yield (defined as the number of emitted photons per unit energy deposited) as good as 2000 ph MeV$^{-1}$, leading to a measured CTR of 136 ps. Ultrafast timing characteristics have been reported by Cs$_3$ZnCl$_5$ with an even shorter decay time of 820 ps and lack of slow decay components[64-65]. Recently, CTR of 180 ps was reported for hot exciton organic scintillators[66]. The scintillator decay time ultimately limits the overall performance of scintillator-based detectors, therefore there is a strong push toward the discovery and development of ultrafast materials that may overcome the limitations of existing technology[8].

In this work, we aim to contribute to this endeavor by investigating the scintillation properties of CsPbCl$_3$ NCs synthesized via a high-throughput room-temperature route and embedded in mass-polymerized polyacrylate composites. To enhance the optical properties and stability towards polymer embedding of as-synthesized CsPbCl$_3$ NCs - which are known to suffer from non-radiative losses and degradation much more severely than their Br- or I-based counterparts[67-69] - we further implemented a post-synthesis surface passivation treatment with CdCl$_2$ directly in a monomeric suspension which lead to near unity photoluminescence (PL) quantum yield ($\Phi_{PL}$), boosted the radioluminescence (RL) and enhanced the NC stability towards the radical initiators used in the living polymerization of the nanocomposite process. The pristine particles, on the other hand, were strongly quenched by the polymerization of the polyacrylate host. Side-by-side temperature controlled PL and RL experiments together with thermally stimulated luminescence (TSL) measurements performed for the first time on this class of LHP NCs revealed the presence of shallow[19] and deep defect states which are completely removed upon treatment. To rationalize the experimental findings, we performed density functional theory (DFT) calculations, confirming the presence of hole deep trap states in pristine CsPbCl$_3$ NCs, arising from superficial undercoordinated chloride ions. These deep trap states can then be successfully eliminated through a Pb to Cd cation substitution, with the latter favoring a more stable tetrahedral coordination at the surface. Radiation hardness studies with very high doses of $^{60}$Co gamma rays reveal exceptional stability of CsPbCl$_3$ NCs to harsh radioactive environments that pair the recently demonstrated performance of the Br-based counterparts[20]. CW and time-resolved RL experiments on polyacrylate nanocomposites containing treated CsPbCl$_3$NCs reveal pure band-edge excitonic scintillation with no defect-related slow tails and an average lifetime as short as 210 ps due to ultrafast multiexciton contributions to the scintillation process, as further confirmed by side-by-side transient transmission (TA) experiments. These results extend our understanding of the scintillation properties of LHP-NCs and make CsPbCl$_3$ a potential valuable candidate for the next generation of scintillator nanocomposites that can furthermore be produced on a large scale.

*Synthesis and resurfacing of CsPbCl$_3$ nanocrystals.* Large scale synthesis of NCs is in general preferable for technological purposes and is essential in the field of radiation detection that requires large quantities of scintillator material for effective interaction with ionizing radiation. Therefore, for the production of CsPbCl$_3$ NCs, we opted for the modified ligand assisted reprecipitation method assisted by turbo-emulsification that we recently introduced for the room temperature synthesis of large batches of CsPbBr$_3$ NCs[22]. Details of the procedure are reported in the **Methods** section. **Figure 1a** (top) shows scanning transmission electron microscopy images acquired with high-angle annular dark field detector (STEM-HAADF) of as-synthesized NCs, showing cubic particles with side lengths of about 9±1 nm and single particle electron diffraction patterns corresponding to the <001> crystal projection of cubic $CsPbCl_3$ (top panel of **Figure 1a**). A high-resolution TEM (HR-TEM) image is also reported highlighting the high crystallinity of the obtained particles despite the room temperature large scale method adopted.

The respective optical properties are reported in **Figure 1b-d**, showing the characteristic sharp absorption peak at 3.05 eV and the narrow excitonic PL at 3.01 eV (413 nm, FWHM= 10 nm). Consistent with previous literature[67-69], as-synthesized $CsPbCl_3$ NCs featured $\Phi_{PL}\sim 3$ % due to severe nonradiative losses that also lead to largely accelerated and multi-exponential decay dynamics, as evidenced by the contour plots of the spectrally resolved PL decay and corresponding decay profiles in **Figure 1c,d**. Various post-synthesis strategies[37, 70-78] are known in the literature for improving the optical properties of LHP-NCs using inorganic salts or organometallic ligands capable of passivating surface vacancies or by inserting isovalent or heterovalent ions into the soft perovskite lattice[79-80]. In order to enhance optical properties of the pristine particles and simultaneously favor their embedding into a polymeric host, we performed a post-synthesis treatment[81] using $CdCl_2$ directly in a lauryl methacrylate (LMA) monomeric solution prior to the free radical polymerization of the PLMA nanocomposite. The choice of PLMA as the matrix was motivated by its non-polar nature and its long alkyl sidechains that create a particularly favorable chemical environment for the NCs[20]. For the treatment, a desired amount of NCs was dispersed in LMA and ethylene glycol di-methacrylate (EGDM) (80:20 %Vol) with a large excess of $CdCl_2$. The solution was stirred vigorously for 3 hours. The unreacted excess of $CdCl_2$ was then decanted and eliminated. The same treatment was also performed directly in toluene to obtain samples of treated NCs that could be deposited on TEM grids for structural/morphological studies. As shown in **Figure 1a** (bottom), no morphological or structural changes were observed after treatment, with the NCs retaining identical shape, size (9 ±1 nm) and crystal structure.

The energy dispersive X-ray spectroscopy (EDX) map and the corresponding elemental composition values are shown in **Supporting Figure S2**, highlighting the presence of Cd ions in concomitance to the NC, in line with the addition of Cd to the NC structure by both substitution of Pb cations as proposed in ref. [81] and/or by filling cation vacancies. In agreement with the substantial invariance of the structural properties upon treatment, both the optical absorption and PL spectra of the NCs remained unchanged (**Figure 1b**). Possibly more importantly, already in monomeric solution, $\Phi_{PL}$ underwent a 30-fold enhancement to $\Phi_{PL}=93\pm6\%$, indicating effective passivation of nonradiative defects. Consistently, the decay dynamics of the pristine particles (**Figure 1c**) turned from strongly multiexponential with lifetime $\tau\sim 250$ ps (measured as the time after which the PL intensity decreased by a factor $e$) to single exponential with lifetime $\tau=2.4$ ns. We highlight that the change in decay time was accompanied by an increase in the zero-delay PL intensity, suggesting the passivation of a manyfold of defect states also responsible for ultrafast quenching faster than the time resolution of our streak camera (ca. 12 ps).

*DFT Simulation of defect passivation.* To elucidate the improvement in the PL efficiency from treating $CsPbCl_3$ NCs with $CdCl_2$ salts, we modeled pristine and treated $CsPbCl_3$ (**Figure 1d, e**), incorporating defects following the work of Bodnarchuk et al. [82] where traps are created upon removal of ionic pairs such as CsCl or $PbCl_2$ from the NC surface. Considering the low $\Phi_{PL}$ of the pristine NCs, we indeed expect the NC surface to be massively decapped. Following this reasoning, the removal of such ionic pairs exposes tetra- (4c-), penta- (5c-), and hexa-coordinated Pb (6c-Pb) atoms at the NC surface.

Some of these ions induce deep trap states due to the orientation of the outermost Cl ions, which experience very low dicoordination (**Figure 1f**), 2c-Cl. These 2c-Cl ions introduce localized intragap states, reducing the optical band gap and quenching the PL.[83] For treated NCs, we hypothesize that the substitution of some Pb surface ions with 4c-Cd atoms eliminates trap states. This optical band gap rectification is attributed to the substitution of cornered 6c-Pb atoms by 4c-Cd, optimizing the position of previously 2c-Cl atoms into three-coordinated Cl (3c-Cl) ions. This adjustment stabilizes the energy of Cl states, shifting them below into the valence states continuum, thereby improving the $\Phi_{PL}$. This mechanism suggests a direct correlation between the coordination geometry at the NC corners and PL efficiency.

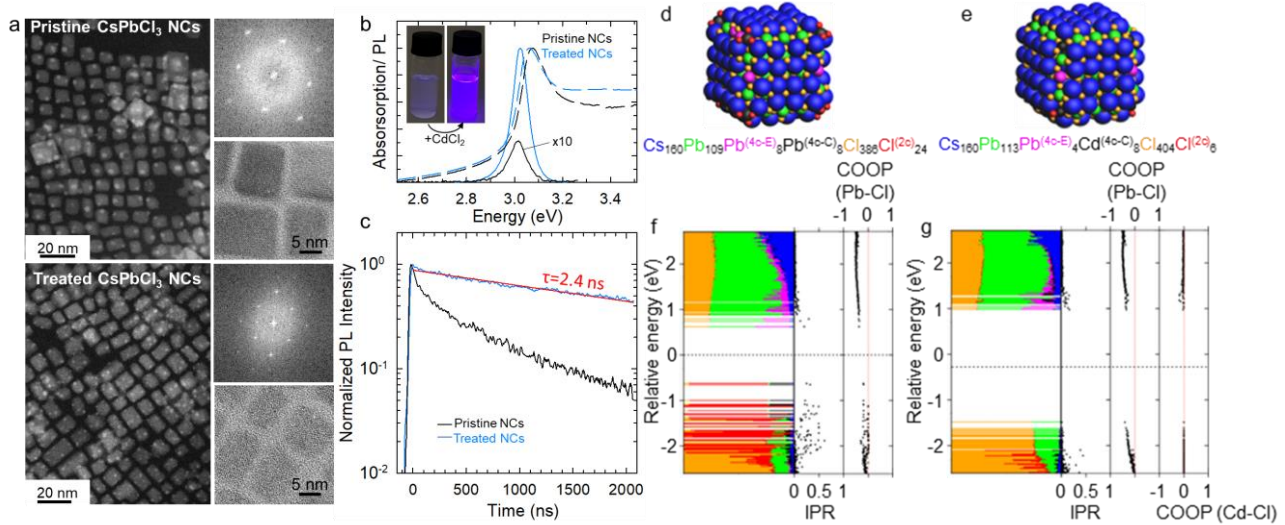

**Figure 1: a)** STEM-HAADF images (left panel) of pristine $CsPbCl_3$ NCs (top) and $CsPbCl_3$ NCs after treatment with $CdCl_2$ (bottom). Top right panel: single particle Fast Fourier Transform (FFT) patterns corresponding to electron diffraction of cubic $CsPbCl_3$ in <001> zone axis. Bottom right panel: HR-TEM images. **b)** Optical absorption (dashed lines) and PL (solid lines, excitation energy 3.35 eV) spectra of pristine (black lines) and $CdCl_2$ treated (blue lines) NCs. For clarity, the PL of pristine NCs has been scaled by 10 times. Inset: photographs of the monomeric solution with the same amount of pristine and treated NCs under UV illumination. **c)** PL decay traces collected at 3.05 eV. The red curve is the single exponential fit of the PL decay of treated NCs. Sketch of the pristine (**d**) and treated (**e**) models with their respective compositions. Density of states (DOS) for the pristine (**f**) and treated (**g**) models. Each horizontal line corresponds to a molecular orbital (MO). The length of the coloured line segments indicates the contribution of each element to a MO. Levels below the dotted line are filled with two electrons; levels above the line are empty. Inverse participation ratio (IPR) and Crystal orbital overlap population (COOP) for the pristine (between Pb and Cl atoms) and treated (between Pb and Cl atoms and Cd and Cl atoms) models are reported next to the respective DOS. For IPR, values close to 0 indicate delocalized MOs, while values close to 1 indicate localized MOs. COOP values less than 0 indicate anti-bonding MOs, values greater than 0 indicate bonding MOs, and values equal to 0 (red vertical line) indicate non-bonding MOs.

*Scintillation properties of $CsPbCl_3$ nanocrystals.* The scintillation properties of the NCs followed a similar fate to the respective PL with $CdCl_2$ treatment, as highlighted by the RL spectra of the untreated and treated NCs measured under the same excitation and collection conditions (**Figure 2a**). The RL spectrum of treated NCs showed a sharp excitonic peak at 3.05 eV, essentially perfectly overlapping with the respective PL, indicating a negligible contribution of shallow defects, which were instead dominant in the RL of the pristine NCs, which was redshifted by 80 meV from the corresponding PL, as systematically observed for $CsPbBr_3$[17, 19] and ascribed to the recombination of excitons trapped in shallow surface defects associated to halide vacancies that are preferentially populated under ionizing excitation. Consistent with the PL data, the RL of treated NCs was over 20 times more intense than for the pristine particles and perfectly resonant to the corresponding PL, thus further confirming the perfect passivation of surface defects. Notably, despite their differences, both samples showed remarkable radiation stability under intense gamma irradiation from a $^{60}$Co source. Specifically, we irradiated the NC composites using the Calliope ISO 9001 certified irradiation facility up to 1 MGy of total absorbed dose and compared the RL intensity to un-irradiated samples. As shown in the **Supporting Figure S3**, both samples retained their scintillation intensity perfectly, indicating exceptional radiation hardness, in agreement with recent data on CsPbBr$_3$ NCs[19]. In comparison, plastic scintillators such as PTP and POPOP in PVT suffered over 80 percent of RL quenching[84-86] upon irradiation with a comparable dose. We have further monitored the RL intensity of CdCl$_2$ treated NCs under continuous X-ray irradiation, showing perfectly stable emission intensity up to 3 kGy dose (**Supporting Figure S4**).

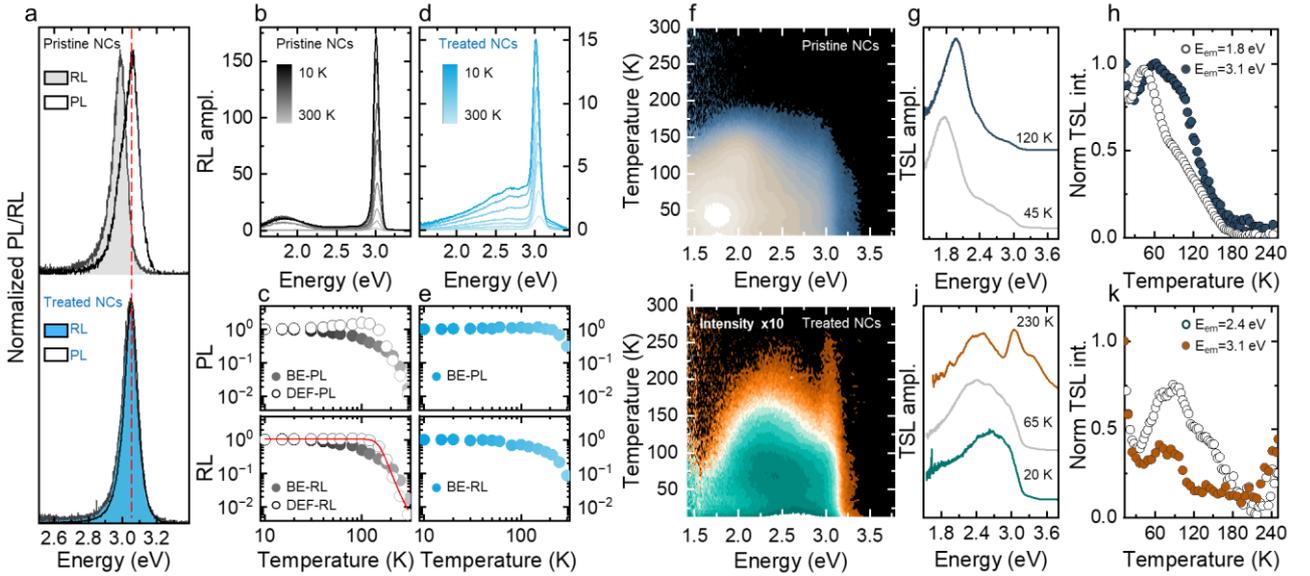

**Figure 2. a)** Room temperature PL (solid line) and RL (shaded area) spectra of pristine (top panel) and treated (bottom panel) CsPbCl$_3$ NCs drop-casted onto a inert PMMA substrate. Dashed line is a guide for the eye indicating the position of the PL peak of pristine NCs. **b)** RL spectra of pristine CsPbCl$_3$ NCs as a function of temperature from 290 to 10 K. **c)** Spectrally integrated PL (top panel) and RL (bottom panel) as a function of temperature for pristine NCs. BE-PL and BE-RL refers to the emission at 3 eV. DEF-PL and DEF-RL refers to the integrated emission in the range 1.55-2.25 eV. Red solid line is the Arrhenius fit with eq 1. **d)** Same as **b)** for treated NCs. **e)** Spectrally integrated PL (top panel) and RL (bottom panel) as a function of temperature for treated NCs. **f)** Contour plot (logarithmic scale) of the spectrally resolved TSL as a function of temperature for pristine CsPbCl$_3$ NCs after X-ray irradiation at 10 K. **g)** TSL spectra extracted from **f)** at 45 K and 120 K. **h)** Glow curve extracted from **f)** at 1.8 eV. **i)** Same as **f)** for treated NCs. **j)** TSL spectra extracted from **i)** at 20 K, 65 K, and 230 K. **k)** Glow curve extracted from **i)** at 2.5 eV and 3.1 eV.

Next, in order to gather deeper insight into the impact of the CdCl$_2$ treatment also on deep intragap defects possibly present in NCs (e.g., due to Pb-related point defects), we performed PL and RL as a function of temperature, which are particularly adapt for probing weakly emitting stable defects[17, 19], as well as TSL experiments. Upon cooling, the RL of pristine CsPbCl$_3$ NCs gradually redshifted and narrowed (**Figure 2b**) due to lattice expansion and reduced phonon coupling[87-89]. At the same time, a defect-related emission at 1.8 eV (indicated as DEF-RL) progressively intensified under both X-ray and UV excitation, reaching a plateau at ~100 K (**Figure 2c,** bottom panel). An Arrhenius fit to the equation, $I(T) = \frac{I_0}{1+A\exp{-\frac{E_A}{k_bT}}}$, yielded an activation barrier for thermal quenching of $E_A$=140 meV; where $I(T)$ is the RL intensity, $I_0$ is the RL intensity approaching 0 K, $A$ is a normalization factor, and $k_b$ is the Boltzmann constant. Interestingly, the band edge RL (BE-RL) showed the same intensification as its corresponding BE-PL upon cooling (**Figure 2c**), suggesting that thermal quenching of the BE luminescence was the main responsible for RL loss at room temperature, and that quenching of the BE and defect emissions occurred via unrelated channels. More importantly, the treated NCs (**Figure 2d, e**) showed no defect-related emission at 1.8 eV, indicating effective passivation of deep defect states. We further notice that the BE-RL peak of the treated NCs was superimposed to a featureless contribution at about 2.5 eV not observed in PL which matched the broad TSL signal observed for both pristine and treated particles, suggesting the presence of additional emissive centers excited exclusively by ionizing radiation that. The absence of such a signal for the pristine NCs suggests that the corresponding trap might participate in the excitation of the main defect peak at 1.8 eV (*vide infra*). We point out that the 3-fold enhancement of the PL of treated NCs upon cooling was due to suppression of thermal losses in the NC solid film (necessary for cryogenic measurements) that featured $\Phi_{PL}$=30%, as commonly observed[90-91]. The picture emerging from the RL measurements as a function of temperature is further enriched by cryogenic TSL experiments, which provide an independent picture of trapping and detrapping also to/from optically dark states that might be responsible for nonradiative quenching of the BE emission. A typical TSL measurement consists of a preliminary X-ray irradiation at low temperature to fill intragap metastable states: the irradiation is successively stopped, and the temperature is gradually increased with a linear heating rate (0.1 K/s), while monitoring the delayed emission due to carrier detrapping. The TSL signal of pristine $CsPbCl_3$ NCs showed no evidence of the BE excitonic peak and came mostly from the defect state emitting at 1.8 eV already observed in RL (**Figure 2f, g**), with a shoulder at ~2.5 eV that was not observed in the low-T RL spectra, suggesting that the corresponding defects might be responsible for populating the lower lying 1.8 eV emitting ones. The related glow curve (i.e. the TSL intensity as a function of temperature) reported in **Figure 2h** showed a TSL peak at 45 K, with a high-temperature shoulder at around 105 K indicating thermal release of deep trapped carriers. After $CdCl_2$ treatment, the TSL intensity drastically decreased (notice the ×10 label in **Figure 2i**) and the BE emission strongly intensified at T>150 K (**Figure 2j**) in agreement with the substantial suppression of surface defects. The corresponding glow curve showed a peak at ~90 K (**Figure 2k**) indicating the complete depletion of the trap states responsible for the weak residual delayed BE luminescence. The broad TSL band at ~2.5 eV was also observed, peaking at T~90 K suggesting the presence of similar defects in both pristine and treated samples, which are also effectively cured by the $CdCl_2$ treatment. Overall, the TSL results confirmed the effective passivation of deep defects arising from the 2c-Cl atoms associated with the octahedrally arranged Pb atoms located at the corners, which are likely responsible for the intragap state dominant in the TSL of pristine $CsPbCl_3$.

*Nanocomposite scintillators based on Cd-treated $CsPbCl_3$ nanocrystals.* Once we clarified the effect of $CdCl_2$ treatment on the optical and scintillation properties of $CsPbCl_3$ NCs, we proceeded with fabricating and testing solid polyacrylic nanocomposites. To this end, 0.33 wt% of 2,2-dimethoxy-2-phenylacetophenone photo-initiator was added to the colloidal suspension of both pristine and treated NCs in LMA. After 15 min under UV irradiation at 365 nm, optical grade PLMA/EGDM (80:20) nanocomposites without macroscopic phase segregation were obtained. The treated NCs showed stable optical properties upon embedding with complete retention (**Figure 3a-c**) of the initial PL spectrum, efficiency, and dynamics. On the other hand, the pristine NCs experienced severe PL quenching, leading to negligible emission efficiency an effective lifetime as short as 110 ps. Based on the promising results for the treated particles, we produced and characterized a set of five nanocomposites containing 0.2 wt%, 0.6 wt%, 0.8 wt%, 3% and 10% of NCs. Because of competitive absorption of the UV curing light by the NCs that inhibited complete polymerization of thick heavily doped polymeric composites, the three low concentration samples were polymerized with thickness of 3 mm, whereas the more heavily doped nanocomposites (3 and 10%) were manufactured as 150μm thick self-standing sheets. The RL spectra are reported in **Figure 3d**, showing concentration independent RL profiles in close match with the peak spectral sensitivity of alkali-based photodetectors commonly used in NA62 experiments (see **Supporting Figure S1**). The scintillation yield was measured by direct comparison in identical conditions with specimen of the plastic scintillator EJ276D of the same thickness as the respective composites (namely 3 mm and 150 μm), showing linearly increasing efficiency up to ~2100 ph/MeV for the most concentrated sample.

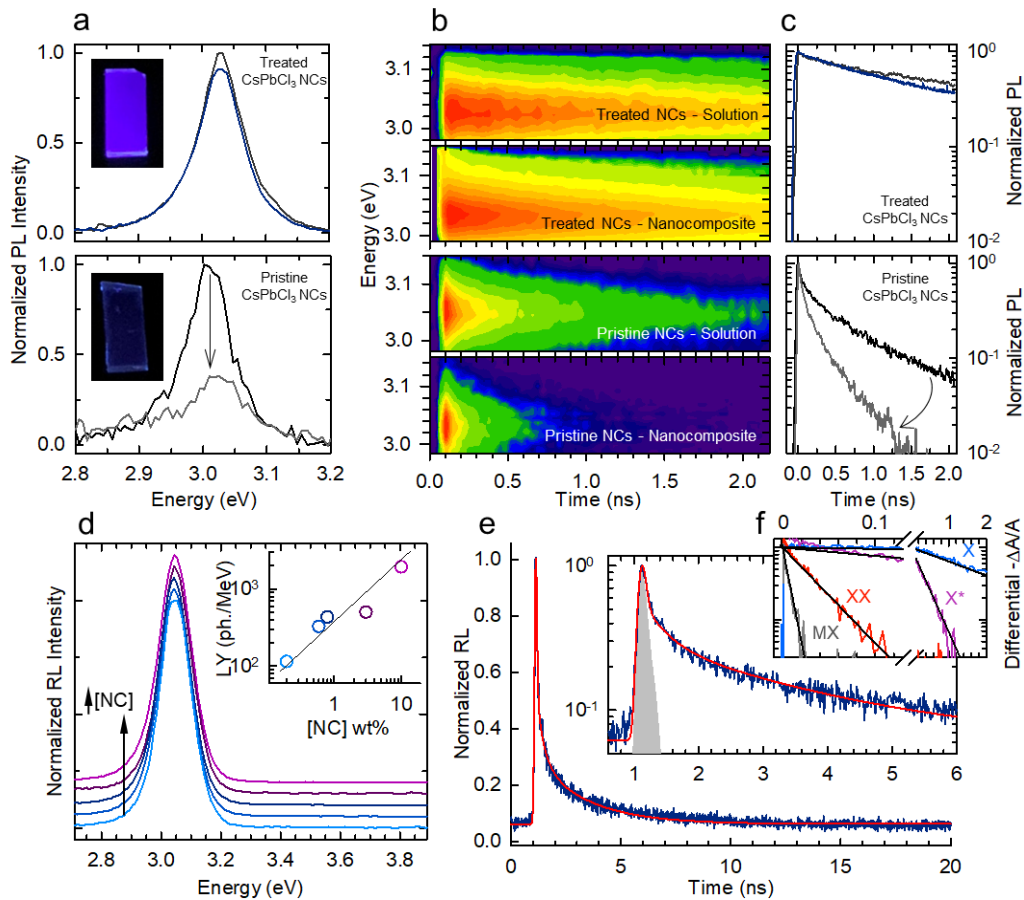

**Figure 3: a)** Normalized PL spectra of Cd-treated $CsPbCl_3$ NCs (top panel) and pristine $CsPbCl_3$ NCs (lower panel) measured in toluene solution side-by-side with the corresponding nanocomposite. **b)** Energy and time-resolved PL decays of pristine and $CdCl_2$-treated $CsPbCl_3$ NCs in LMA before and after polymerization of the nanocomposite matrix ([NC]=0.8wt%). **c)** PL decay traces extracted from 'b' together with the corresponding toluene solutions. Inset: photographs of the final nanocomposites embedding the same amount of pristine and Cd-treated $CsPbCl_3$ NCs under UV light. **d)** Normalized RL spectra of Cd-treated $CsPbCl_3$ nanocomposites at increasing NC loading concentrations. The spectra have been vertically shifted for clarity. Inset: measured light yield values for the nanocomposites. The same color scheme applies throughout the panel. **e)** Time-resolved RL decay of the 0.8 wt% loaded nanocomposite. The scintillation decay is shown in a linear scale over 20 ns. The inset shows details of the ultrafast component (over 6 ns) in semilogarithmic scale; the shaded grey line represents the system IRF. The red curves are the fit function. **f)** Differential TA curves extracted from the bleaching signal of the band-edge excitonic states at increasing pump fluences, representing single (X), charged (X*), bi (XX), and higher order multiexciton (MX) dynamics. The black lines are the fit with single exponential decay functions.

Next, we interrogated the timing performance of our CsPbCl$_3$ NCs-based nanocomposites by performing scintillation kinetics measurements in time-correlated single photon counting mode under X-ray excitation. The scintillation decay curve of a representative nanocomposite ([NC] = 0.2 %wt) is shown in **Figure 3e** together with the instrument response function (IRF, FWHM = 120 ps). Importantly, in agreement with very recent results on CsPbBr$_3$ NCs[12, 17, 20, 92-93], the RL decay trace showed a substantial ultrafast contribution ascribed to the recombination of multi-excitons generated upon X-ray excitation, followed by a slower tail ascribed to the decay of charged and neutral excitons, as highlighted by TA measurements (*vide infra*). To quantify the timing contributions, we fitted the experimental decay curve with a convolution of the instrumental response of our detection chain and a triple exponential function accounting for the decay of single, charged, and double excitons and calculated the effective scintillation lifetime ($\tau_{EFF}$) as the harmonic average of scintillation components weighted by their respective time-integrated relative contributions. The resulting best fit corresponding to a biexciton lifetime of 30 ps (weight 20%) followed by a 320 ps charged exciton decay (10%) and a single exciton lifetime of 2.3 ns (70%) - in good agreement with both the time-resolved PL data and the TA dynamics - is shown as a red line in **Figure 3e**. In order not to over-interpret the biexciton contribution that is faster than the instrument response function, we also performed the fitting procedure by forcing the biexciton lifetime up to 60 ps, close to the fall time of the response function (~75 ps, see **Figure S5** and **Table S1**). The resulting $\tau_{EFF}$ ranged between 210 ps and 300 ps which is substantially faster than the typical values measured for Br-based lead halide NCs [20, 94], UV-emitting RE-doped crystal scintillators[95-97], as well as the sub-ns decay time of the core-valence transition in cross luminescent scintillators. Such ultrafast timing capability therefore has great potential relevance for time-tagging technologies. To evaluate the timing resolution reachable with our nanocomposites in TOF-PET, we calculated the $CTR = 3.33\sqrt{(\tau_{RISE} \times \tau_{EFF})/N}$ , where $\tau_{RISE}$ is the 10-90% signal risetime (due to the detection chain and set to 100 ps) and $N$ is the estimated number of emitted photons for a 511 keV excitation for the most concentrated nanocomposite. We reached an estimate CTR=14.7-18.2 ps which is very promising for breaking the 10 ps limit in fast timing applications. Crucially, the potential to increase the scintillation yield of ultrafast scintillating NCs to achieve ultimate performance makes them valid candidates to overcome the limitations of current technologies.

Finally, to confirm the ascription of the various kinetic components to the decay of excitonic species created under high-energy excitation, we performed ultrafast TA measurements. The normalized TA spectra and the corresponding dynamics as a function of the pump fluence are reported in **Supporting Figure S6** showing the progressive emergence of faster components upon increasing the average excitonic population, ⟨N⟩. The detailed analysis of TA dynamics is shown in **Figure 3f** by extracting the components of each excitonic order by sequential subtraction of the lower ⟨N⟩ contributions[98-99]. In the single exciton regime (indicated with X, corresponding to an average exciton occupancy ⟨N⟩≪1), the TA spectrum is due to the bleach signal of the 1S band edge transition at 3.05 eV at all times, with a bleach recovery lifetime $\tau_X$ = 2.3 ns resembling the PL lifetime measured at vanishingly low excitation power and matching well the slowest contribution to the RL kinetics. At progressively higher pump fluences, following the statistical state filling of the excitonic levels, the TA dynamics developed the ultrafast decay components characteristic of biexcitons ($\tau_{XX}$ = 33 ps) and higher-order excitons ($\tau_{MX}$ = 6.8 ps), together with an intermediate contribution ($\tau_{X^*}$ ~ 300 ps) ascribed to the recombination of photo-charged excitons, as previously observed in nanocomposites with $CsPbBr_3$ NCs[17, 20].

In conclusion, we have fabricated and characterized for the first time the scintillation properties of $CsPbCl_3$ NCs both as synthesized and after a defect passivation treatment with $CdCl_2$. The resurfacing treatment, performed directly in the monomeric solution prior to polymerization, massively enhanced the optical properties by suppressing Pb-related defects and stabilized the NCs towards the living radical polymerization of the polyacrylate host, which preserves the optical properties of the NCs. DFT calculations, aligned with experimental measurements, demonstrate that hole deep trap states in $CsPbCl_3$ NCs, primarily arising from undercoordinated chloride ions, 2c-Cl, are effectively eliminated by substituting Pb with Cd. This substitution optimizes the coordination of Cl atoms, thereby stabilizing their energy states below the valence band edge and eliminating the band gap from intragap states. Consequently, a direct correlation between the coordination geometry of the more exposed regions of the NCs (corners) and the emission yield can be established. Steady-state and time-resolved RL experiments reveal a purely excitonic scintillation of the treated particles, with no spurious slow intragap contributions, and ultrafast scintillation due to contributions from single and multi-excitonic species. Our results fill the knowledge gap in the understanding of nanoscintillators based on lead halide perovskite NCs and provide a potential route to obtain nanocomposite scintillators emitting in the UV-blue spectral range that match well the spectral response of ultrafast photodetectors largely employed in scintillation applications and provide a valuable alternative for effective sensitization of secondary emitters in the visible spectral range.

## Supporting Information

Experimental methods, computational details, energy dispersive X-ray spectroscopy (EDX) analysis, radiation hardness study, fitting procedure, transient transmission data.

## Acknowledgments

This work was funded by Horizon Europe EIC Pathfinder program through project 101098649 – UNICORN, by the PRIN program of the Italian Ministry of University and Research (IRONSIDE project), by the European Union—NextGenerationEU through the Italian Ministry of University and Research under PNRR—M4C2-I1.3 Project PE_00000019 "HEAL ITALIA", by European Union's Horizon 2020 Research and Innovation programme under Grant Agreement No 101004761 (AIDAINNOVA). This research is funded and supervised by the Italian Space Agency (Agenzia Spaziale Italiana, ASI) in the framework of the Research Day "Giornate della Ricerca Spaziale" initiative through the contract ASI N. 2023-4-U.0

# Supporting Information

**Methods**

*Chemicals:* Cesium carbonate (Cs$_2$CO$_3$, 99%), Lead (II) chloride (PbCl$_2$, 99.99%) were purchased from Fluorochem. Propionic acid (PA, >99.5%). Tetrabutylammonium bromide (TBAB, >98%), methyl methacrylate (MMA, 98%), lauryl methacrylate (LMA, 98%), Ethylene glycol dimethacrylate (EGDM, 97.5%), 2,2-dimethoxy-2- phenylacetophenone (Irgacure 651, 99%) and cadmium chloride (CdCl$_2$, 99,99%) were purchased from Sigma-Aldrich. Oleylamine (OLAM, 90%) was purchased from Acros Organic. Isopropanol (HiPerSolv chromanorm for HPLC, >98%), Heptane (Gpr rectapur, 99.8%) were purchased from VWR. Turbo emulsifier homogenizer was bought from IKA and it is composed of motor group T25/T50 digital ultra turrax and dispersing tool with code S25N–25G and S50N-G45M for volumes respectively until 2 L and above.

*Synthesis of CsPbCl$_3$ NCs.* Solution A is prepared dissolving Cs$_2$CO$_3$ (97.5 mg, 0.3 mmol) in propionic acid (0.3 mL, 3.99 mmol) and then and diluting with 180 mL of heptane/isopropanol (2:1 V/V). Solution B is obtained by dissolving PbCl$_2$ (834 mg, 3 mmol) with magnetic stirring in a mixture of oleylamine (8.9 mL, 27 mmol), propionic acid (2 mL, 27 mmol) and isopropanol (2 mL) at 80 °C. After complete dissolution of the precursors, the mixture is allowed to cool to room temperature. Solution A is then stirred into a 250 mL beaker with a turbo-emulsifier homogenizer (15k rpm) and Solution B is quickly added. The mixture is allowed to develop under homogenization for 30 seconds. Stirring is stopped and 60 mL of isopropanol is added to the crude solution to precipitate the NCs. The whole process takes about 24 hours. After this time, the product is collected by centrifugation at 4500 rpm for 2 min. The supernatant is discarded and the precipitate dried, weighed (210 mg) and finally stored in solid state under argon atmosphere.

*Cadmium chloride treatment of CsPbCl$_3$ NCs.* Pristine CsPbCl$_3$ NCs (4 mg) are dissolved in a 8 mL mixture of dispersed in LMA and EGDM (80:20 %Vol). Subsequently, an excess of CdCl$_2$ in powder (100 mg) is added and the prepared solution is vigorously stirred for 3 hours while PL is constantly monitored to follow the evolution of NCs upon the treatment. Once the treatment is completed (revealed by the end of the NCs PL increase), the stirring is stopped, the solution is mildly centrifugated (500 rpm 1 minutes) and the clear solution containing Cd-treated NCs is separated from the precipitated excess of CdCl$_2$. The same treatment was also performed directly in toluene to obtain samples of Cd-treated NCs that could be deposited on TEM grids for structural/morphological studies.

*Fabrication of polymer nanocomposites.* The polymerization is performed directly on the same solution which underwent the Cd-treatment. 2,2-dimethoxy-2-phenylacetophenone photo-initiator (0.33 wt%) was added to the colloidal suspension of NCs in LMA/EGDM. The mixture is then transferred into a sealed mold formed by two plain glass slabs divided by a silicon gasket and is then transferred into a polymerization chamber with continuous 365 nm light After 15 min under UV irradiation, optical grade nanocomposites without macroscopic phase segregation were obtained.

*Transmission electron microscopy.* High-Angle Annular Dark Field (HAADF) images in scanning TEM (STEM) mode and the high-resolution TEM images (HRTEM) of powdered NCs were acquired in a JEOL JEM- 2200FS transmission electron microscope equipped with in-column Omega filter, operating at 200 kV and coupled with an EDX Oxford Xplore silicon-drift detector, with 80 mm$^2$ effective area.

*Optical spectroscopy measurements.* Optical absorption spectra were measured in toluene with a Cary 50 UV–Vis spectrophotometer. The PL spectra were excited using a laser source at 3.50 eV (355 nm); the emitted light was collected using a custom apparatus featuring a liquid nitrogen-cooled, back-illuminated, and UV-enhanced charge-coupled device detector (Jobin-Yvon Symphony II) coupled to a monochromator (Jobin-Yvon Triax 180) with 100 lines/mm gratings and corrected for the spectral response of the system. Temperature-controlled PL measurements were performed in the 10-300 K range by lowering the temperature with a closed-cycle He cryostat and controlling the temperature with a Lakeshore 330 temperature controller. Time-resolved PL measurements were carried out using frequency-doubled 250-fs pulses from a 78 MHz Ti:Sapphire laser at 3.2 eV. The emitted light was collected with a Hamamatsu streak camera (time resolution ~10 ps). Ultrafast transient absorption spectroscopy measurements were performed on Ultrafast Systems' Helios TA spectrometer. The laser source was a 10 W Ytterbium amplified laser operated at 1.875 kHz producing ~260 fs pulses at 1030 nm and coupled with an independently tunable optical parametric amplifier from the same supplier that produced the excitation pulses at 3.5 eV. After passing the pump beam through a synchronous chopper phase-locked to the pulse train (0.938 kHz, blocking every other pump pulse), the pump fluence on the sample was modulated from 1.3 µJ cm$^{-2}$ to 170 µJ cm$^{-2}$. The probe beam was a white light supercontinuum.

*Radioluminescence measurements.* The excitation source was unfiltered X-ray radiation generated by a Philips PW2274 X-ray tube, with a tungsten target, equipped with a beryllium window and operated at 20 kV. At this operating voltage, a continuous X-ray spectrum is produced by *Bremsstrahlung* mechanism superimposed to the L and M transition lines of tungsten, due to the impact of electrons generated through thermionic effect and accelerated onto a tungsten target. The RL was collected using the same custom apparatus used for the PL measurements. The light yield values were evaluated using a relative method by direct comparison of the RL response in the same experimental conditions of Cd-treated CsPbCl$_3$ nanocomposites and commercial EJ276D plastic scintillator of the same size used as reference, with a LY of 8600 photons/MeV. Cryogenic RL measurements are performed in the 10−300 K interval by lowering the temperature with a closed-cycle He cryostat and controlling the temperature with a Lakeshore 330 temperature controller.

*TSL and AG measurements.* Wavelength-resolved TSL measurements at cryogenic temperatures are carried out by using the same detection system as for RL measurements on the same pristine and Cd-treated CsPbCl$_3$ NCs drop-casted onto an inert polymethylmethacrylate slab used for the RL experiments vs *T*. Cryogenic TSL measurements are performed in the 10−300 K interval, with a heating rate of 0.1 K·s$^{-1}$ after X-ray irradiation at 10 K up to a dose of 5 Gy. The dose values for X-ray irradiation were obtained with a calibrated ionization chamber and evaluated in air. The shape of the TSL signal vs *T* (the so-called glow curve) has been corrected for the variation of the RL emission intensity vs *T*, to decouple the trap contribution to the emission from the other mechanisms involved in the scintillation process. Wavelength-resolved AG measurements are carried out by monitoring the luminescence emission at a constant temperature ($T = 10$ K) as a function of delay time after the suppression of X-ray irradiation: the samples are irradiated with the same doses used for the TSL experiments.

*Time-resolved scintillation experiments.* Samples were excited with a Hamamatsu XRT N5084 pulsed tungsten X-ray tube operating at 40 kV, where a PicoQuant PDL 800-B pulsed diode laser with 40 ps pulse width (full-width-at-half-maximum - FWHM) acts as excitation source of the X-ray tube. The energy spectrum of the produced X-rays ranges from 0 to 40 keV with a pronounced peak between 9 and 10 keV, due to tungsten L-characteristic X-ray, and mean energy of about 15 keV. The X-rays hit the sample after crossing a brass collimator. The scintillation light is collected in TCSPC by a Becker & Hickl HPM 100-07 hybrid photomultiplier tube (HPM). The signal of the HPM was processed by an ORTEC 9237 amplifier and timing discriminator and acted as stop signal at a Cronologic xTDC4 time-to-digital-converter (TDC). The start signal was given by the external trigger of the pulsed laser. The overall impulse response function (IRF) of the system was obtained as the analytical convolution between the measured IRF of the laser together with HPM and the IRF of the X-ray tube, resulting in around 160 ps FWHM. The RL was spectrally selected using an optical bandpass filter at 500 nm with 40 nm FWHM mounted to the HPM that removed parasitic contributions due to air excitation by X-rays.

*γ-ray irradiation experiments.* CsPbCl$_3$ NCs were placed in a polypropylene sealed vial whose gamma ray attenuation is negligible (Eppendorf Tubes). The vial was irradiated in a pool-type gamma irradiation chamber equipped with a $^{60}$Co (mean energy ~1.25 MeV) γ-source rods array, uniformly irradiating the NCs at 3.05 kGy$_{air}$/h dose rate value up to 1 MGy. The irradiation has been carried out at the CALLIOPE Gamma Irradiation Facility at ENEA Casaccia Research Centre (Rome). No significant γ-ray attenuating material was interposed between the source and the sample. Throughout the paper, the given dose is in air.

*Computational details.* Geometry optimizations were carried out at the DFT level with a PBE exchange-correlation functional[1] and double-ζ basis set, as implemented in the CP2K quantum chemistry software package[2]. Relativistic effects were considered using effective core potentials. The IPR and COOP were computed as in the main reference[83], using the workflows implemented in the Nano-QMFlows package[3]. To simplify the analysis, the size of our nanocrystal models is 3 nm of edge length. The composition of each nanocrystal is kept in such a way that the overall system is charge neutral and charge balanced, with each of the ions in their most stable oxidation state, i.e. Cs as +1, Pb and Cd as 2+, Cl as -1.

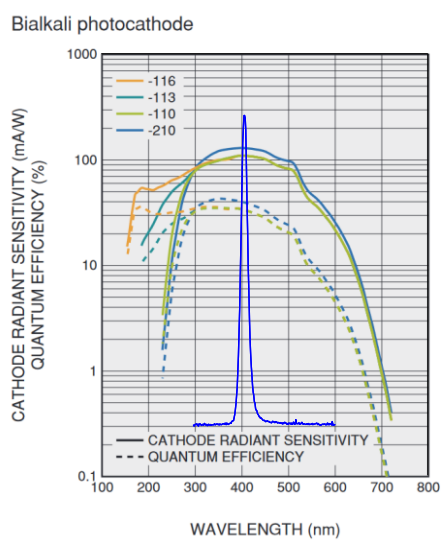

**Figure S1**: Typical spectral response of bialkali photocathode of Hamamatsu R9880U Photomultiplier tubes used in the NA62 experiment to study rare kaon decays (https://www.hamamatsu.com/eu/en/product/optical-sensors/pmt/pmt_tube-alone/metal-package-type/R9880U-210.html; accessed 11-April-2024) together with the RL spectrum of our $CsPbCl_3$ NCs based nanocomposites.

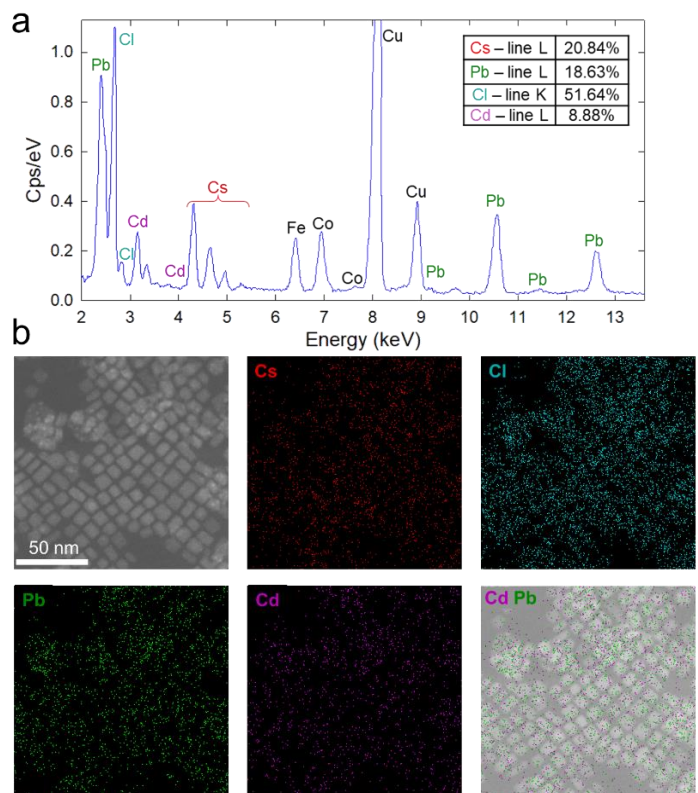

**Figure S2.** Energy dispersive X-ray spectroscopy (EDX) map (a) and the corresponding elemental composition (b). Cu, co and Fe signals are given by the sample holder.

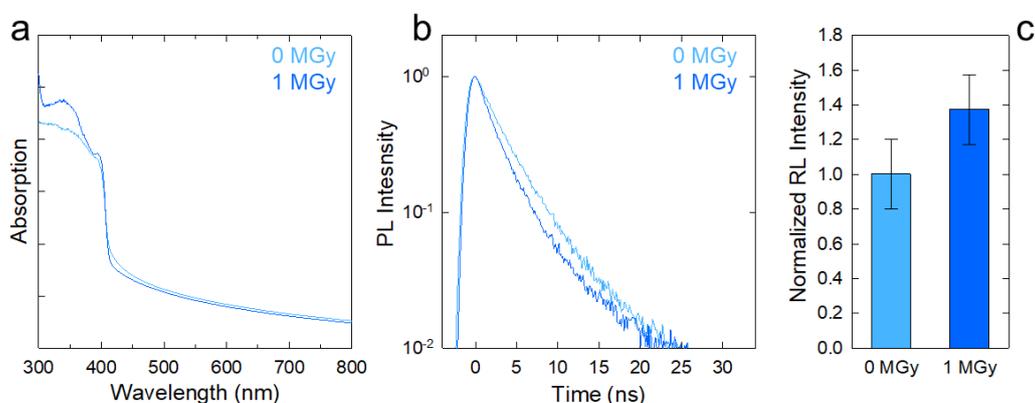

**Figure S3.** Absorption spectra (a), PL decay traces (b) and normalized RL intensity of CdCl$_2$-treated CsPbCl$_3$ PNC embedded into a PLMA-EGDM nanocomposite before (light blue curves/bars) and after (dark blue curves/bars) exposition to 1 MGy γ-ray dose using a $^{60}$Co source. The more pronounced absorption in the UV of the sample exposed to 1 MGy is given by the partial degradation of the PLMA-EDGM matrix.

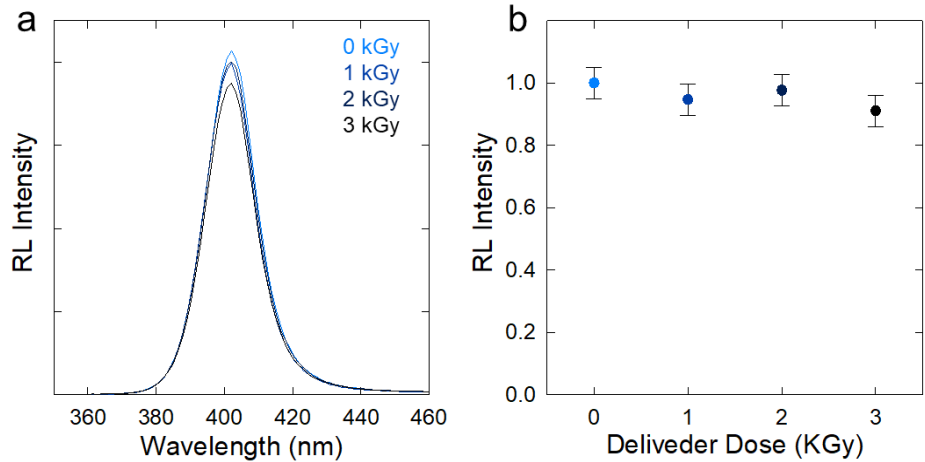

**Figure S4.** RL **a.** spectra of CdCl$_2$-treated CsPbCl$_3$ PNC embedded into a PLMA-EGDM nanocomposite during continuous exposition to X-rays. The corresponding integrated intensities vs. X does are shown in **b**.

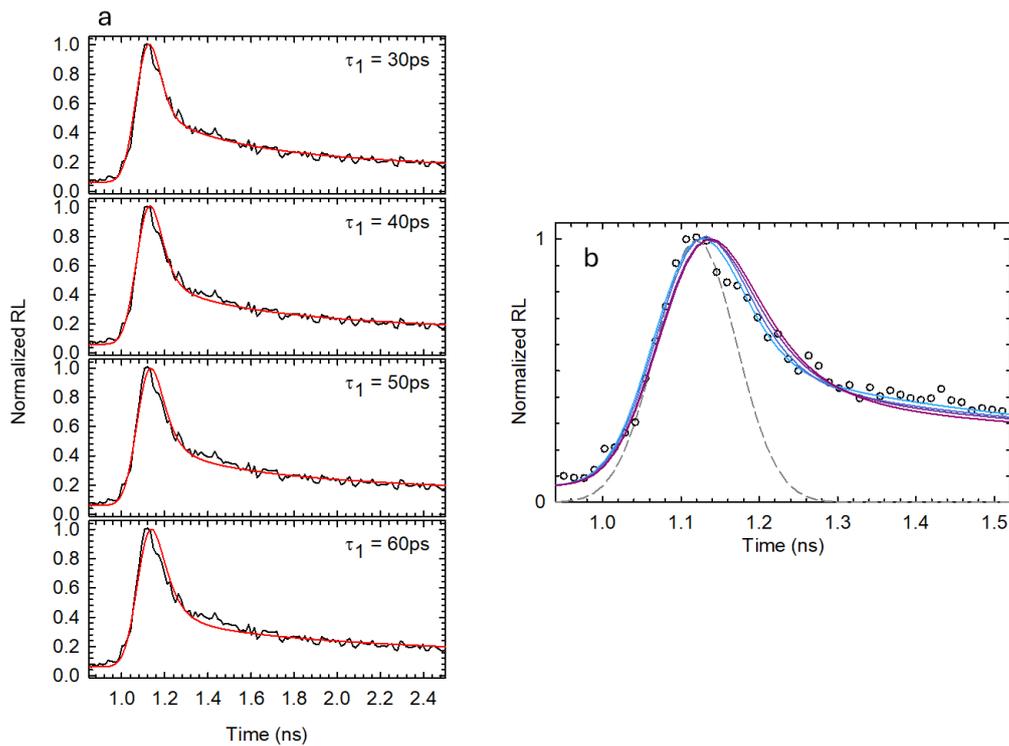

**Figure S5. a.** RL decay curves of the 0.8 wt% loaded nanocomposite fitted with different parameters as shown in Table S1. **b.** Detail of the ultrafast scintillation component over the first ~500 ps. The solid lines represent the fits with a tri-exponential decay function convolved with the instrument response function (dashed line), where the initial ultrafast component was left as a free parameter (resulting in $\tau_1$ = 30 ps) or forced to $\tau_1$ = 40, 50, 60 ps.

| Cd-treated CsPbCl$_3$ [NC] 0.8 wt% | $\tau_1$ (ps) | $w_1$ (%) | $\tau_2$ (ps) | $w_2$ (%) | $\tau_3$ (ps) | $w_3$ (%) | $\tau_{eff}$ (ps) | CTR Estimated (ps) |
|---|---|---|---|---|---|---|---|---|
| Best fit | 30 | 12 | 320 | 13 | 2300 | 75 | 210 | 14.7 |
| With $\tau_1$ = 40 ps | 40 | 13 | 320 | 10 | 2300 | 77 | 250 | 16.1 |
| = 50 ps | 50 | 14 | 320 | 9 | 2300 | 77 | 295 | 17.4 |
| = 60 ps | 60 | 15 | 320 | 6 | 2300 | 79 | 320 | 18.2 |

**Table S1.** Fit parameters (decay time and relative integrated weight, $w_i$) used to analyze the time-resolved RL intensity decay spectra recorded on CsPbCl$_3$ PLMA nanocomposite. The RL dynamics were analyzed in a least squares sense using a triple exponential function convolved with the instrument response function. The effective decay time ($\tau_{eff}$) was calculated using the renormalized ratio of all the components according to $\tau_{eff}^{-1} = \sum_{i=1}^{3} R_i/\tau_i$, $R_i = w_i/(w_1 + w_2 + w_3)$. The CTR has been estimated using the formula $CTR = 3.33\sqrt{\frac{\tau_{rise} \cdot \tau_{eff}}{N}}$, where the rise time ($\tau_{rise}$) was set to 100 ps, and the estimated number of photons emitted for a 511 keV excitation was set to N = 1073.

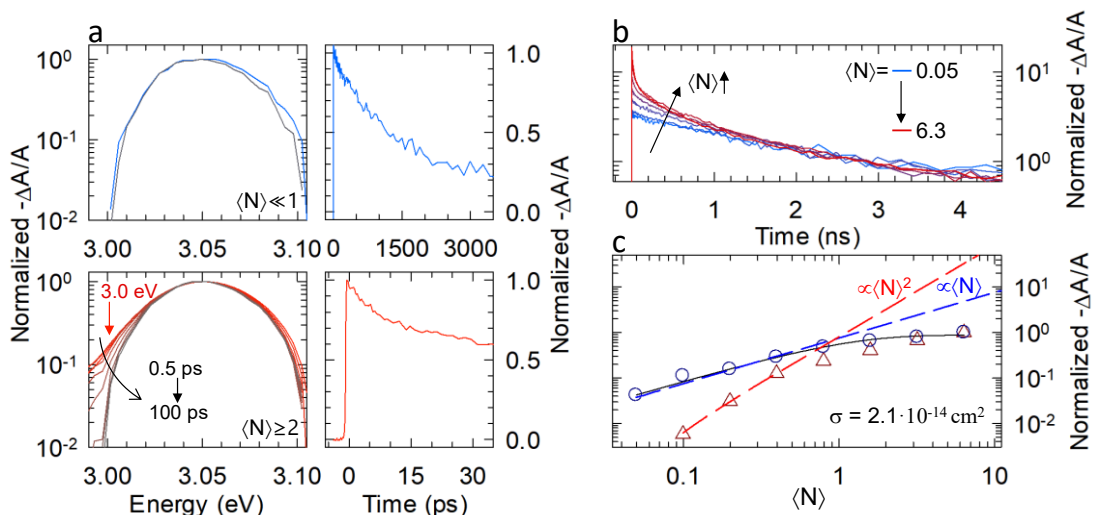

**Figure S6. a)** Normalized transient absorption spectra at increasing delay time for the [NC] 0.2 wt% nanocomposite embedding tread CsPbCl$_3$ NCs after the excitation pulse for single-excitonic regime ($\langle N \rangle \ll 1$, upper panel), and at higher average exciton population ($\langle N \rangle \geq 2$, lower panel) showing the emergence of bi-spectral contributions on the low energy side of the TA spectrum. The respective decay traces, taken at 3.05 eV and 3.00 eV, are reported in the right-hand panels showing the single-exciton ($\tau_X$ = 2.3 ns) and the biexciton ($\tau_{XX}$ = 33 ps) decay rates discussed along the main text. **b)** TA dynamics at increasing average exciton population $\langle N \rangle$. The TA traces have been normalized over the single excitonic long-lived tail (between 2-3 ns) to emphasize the emergence of faster components upon increasing the excitation fluence. **c)** Pump-fluence dependence of the single exciton TA amplitude (black circles) evaluated from the late-time TA amplitude (~3 ns), when there are virtually no contributions from short-lived multiexcitons and each photoexcited NC is populated with a single exciton independent of its initial occupancy. This is further corroborated by the linear growth rate with $\langle N \rangle$ (dashed blue line) before single exciton state filling occurs when $\langle N \rangle$ approaches $\langle N \rangle$=1. The amplitude of the multi-excitonic component (red triangles) is obtained by subtracting the TA amplitude relative to single excitons from the total TA intensities at early times following the method introduced in Supporting Ref.4. At low pump fluences ($\langle N \rangle$ < 0.4), the multi-excitonic component follows a quadratic scaling with $\langle N \rangle^2$, as expected for biexcitons (dashed red line). The solid line represents the fit assuming a Poisson statistic of the NC occupancy from which the absorption cross section ($\sigma$ = 2.1·10$^{-14}$ cm$^2$) at the pumping wavelength was obtained.